\documentclass[aps,prb,twocolumn,floatfix,showpacs]{revtex4-1}
\usepackage{amsmath,amssymb,graphicx,bm}
\begin{document}
\title{Universality of vertex corrections to the electrical
  conductivity in models with elastically scattered electrons}

\author{V.  Jani\v{s}} \author{V. Pokorn\'y}

\affiliation{Institute of Physics, Academy of Sciences of the Czech
  Republic, Na Slovance 2, CZ-18221 Praha, Czech Republic }
\email{janis@fzu.cz}

\date{\today}


\begin{abstract}
  We study quantum coherence of elastically scattered lattice
  fermions. We calculate vertex corrections to the electrical
  conductivity of electrons scattered either on thermally equilibrated
  or statically distributed random impurities. We demonstrate that the
  sign of the vertex corrections to the Drude conductivity is in both
  cases negative. Quantum coherence due to elastic back-scatterings
  always leads to diminution of diffusion.
\end{abstract}
 \maketitle 
\section{Introduction}\label{sec:Intro}

It is very difficult to describe full electron correlations due to a
number of complex phenomena related to the quantum character of the
electron. It is hence appropriate to approach the full description of
electron correlations iteratively in several stages. The first one is
the static mean-field approximation of the Hartree type. Such a
mean-field theory completely neglects both charge and spin
fluctuations and reduces the interacting system to a Fermi gas with
renormalized, self-consistently determined, densities. Such a
simplification may deliver reliable results only for macroscopic
static quantities in the weak-coupling limit. Electron correlations in
this approximation have no impact on dynamical and transport
properties.

The next level in a comprehensive modeling of strong electron
correlations are models and approximations allowing for charge
fluctuations. In these models the spin of electrons does not play a
significant role and electrons are subjected only to a potential
scattering. It means that electrons are only scattered on fluctuations
of the atomic potential in the lattice. The potential fluctuations are
caused by impurities that may be distributed in the crystal either
regularly or randomly. The paradigm for the former situation is the
Falicov-Kimball model\cite{Falicov69} (FKM) and for the latter the
Anderson model of disordered electrons (DAM).\cite{Anderson58} Unlike
the static mean-field approximations, the models with a potential
scattering lead to quantum dynamical effects and are applicable to the
entire range of the interaction strength (variance of the potential
fluctuations). The potential scattering does consequently affect
spectral and transport properties of the system.

A common feature of the models with potential scatterings only is that
energy is conserved during scattering events and need not be treated
as a dynamical variable. Each energy, however, is renormalized in a
different manner and hence the energy (frequency) is used as an
external label. Conservation of energy in scattering events is a
significant simplification in the description of electron
correlations. It allows for an exact solution in the limit of infinite
spatial dimensions (dynamical mean-field theory), where the effect of
strong potential fluctuations may be studied without uncontrolled
approximations.\cite{Janis89,Brandt89,Janis90,Janis92b}

The two models, FKM and DAM, are standardly used for different
purposes. The former one is aimed at a description of quantum
fluctuations caused by electron correlations in thermally equilibrated
states, while the latter one was introduced so that a response of a
disordered electron gas to weak electromagnetic non-equilibrium
perturbations can be estimated in a controlled way.  Both the models
have served well their original purposes. The Falicov-Kimball model
has been successfully applied to a simplified description of
correlation-induced metal-insulator\cite{Ramirez70,Freericks92} and
valence-change\cite{Felner86} transitions in rare-earth compounds, or
atoms in optical lattices.\cite{Ates05, Farkasovsky08} The disordered
Anderson model has been used to describe the spectral and transport
properties of metallic alloys\cite{Gonis92} and vanishing of
diffusion, called Anderson localization.\cite{Kramer93} There have
been efforts to describe the combined effect of electron correlations
and randomness in the disordered Falicov-Kimball model\cite{Janis92b}
or Anderson localization in FKM.\cite{Byczuk05} Only a few attempts
have, however, been made in the calculation of the response of FKM to
non-equilibrium perturbations beyond the mean-field
approach.\cite{Freericks03} In particular, it is little known about
the electrical conductivity of FKM beyond the mean-field, Drude
contribution.\cite{deVries93}

It is the aim of this paper to fill up this gap and to propose a
systematic way how to calculate vertex corrections to the Drude
(mean-field) electrical conductivity in models with elastic
scatterings only, that is FKM, DAM or a disordered FKM. The method we
use is an expansion around the mean-field solution obtained from the
asymptotic limit to high spatial dimensions. We calculate the
leading-order vertex correction in high spatial dimensions being of
order $1/d^2$, while the Drude conductivity is of order $1/d$.  We
demonstrate that the vertex corrections have a universal behavior for
all models of elastically scattered electrons and are always negative,
independent of whether they are caused by an external random potential
(quenched randomness) or by static, thermally equilibrated electron
correlations (annealed randomness).

\section{Electrical conductivity of elastically scattered electrons}
\label{sec:Conductivity-definitions}

Elastic scatterings of electrons are caused by either internal or
external fluctuations of atomic potentials of frozen ions forming a
crystalline lattice. That is, interactions of electrons of the same
sort are excluded in models with elastic scatterings. They are
actually forbidden in spinless models with locally interacting
fermions.  These models hence contain either more than one type of
electrons or an externally governed distribution of atomic
potentials. We consider only the elementary version of such models and
take into account only two types of electrons, extended and localized
ones. We further assume homogeneity in the distribution of the
localized electrons and hence the non-interacting part of the
Hamiltonian describing such a situation reads
\begin{subequations}\label{eq:FK-Hamiltonian}
  \begin{align}\label{eq:FK-H0}
    \widehat{H}_0 &= \sum_{\mathbf{k}}\epsilon(\mathbf{k})
    c^{\dagger}(\mathbf {k}) c(\mathbf{ k}) + E_f \sum_i f^{\dagger}_i
    f^{\phantom{\dagger}}_i \ .
  \end{align}%
  We are interested in dynamical properties of the delocalized
  electrons induced by fluctuations of the atomic potential the
  extended electrons feel. To this purpose we introduce an interacting
  term
  \begin{align} \label{eq:FK-HI} \widehat{H}_I & = \sum_i \left[V_i +
      U f^{\dagger}_i f^{\phantom{\dagger}}_i \right] c^{\dagger}_{i}
    c^{\phantom{\dagger}}_{i}
  \end{align}\end{subequations}
where $c_i = N^{-1}\sum_{\mathbf{k}}c(\mathbf{k}) \exp
\{-i\mathbf{k}\cdot \mathbf{R}_i\}$. We denoted $V_i$ atomic levels of
the ion situated in the elementary cell centered around the lattice
vector $\mathbf{R}_i$, and $U$ is the interaction strength between the
extended and localized electrons. We generally assume that the atomic
potential $V_i$ is a random variable with a static site-independent
probability distribution of its values determined externally.  If
$U=0$ the full Hamiltonian $\widehat{H} = \widehat{H}_0 +
\widehat{H}_I$ is that of the Anderson model with disordered electrons
and if $V_i=0$, the full Hamiltonian describes the Falicov-Kimball
model. We hence see that the most general elementary Hamiltonian for
elastically scattered electrons is just a disordered FKM. Actually,
both the contributions to the interacting Hamiltonian from
Eq.~\eqref{eq:FK-HI} introduce randomness into the distribution of
atomic potentials the extended electrons feel. Potential $V_i$
represents a static (quenched) randomness and the interaction $U$ a
dynamical (annealed) one. Both contributions can be treated on the
same footing.

\begin{figure}
  \includegraphics[scale=1]{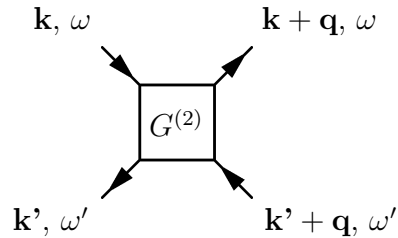}
  \caption{Translationally invariant (averaged) two-particle Green
    function in momentum space in the notation used in the Kubo
    formula for electrical
    conductivity. }  \label{fig:G2-Notation} \end{figure}

We are not interested in this paper in equilibrium thermodynamic
properties of FKM but rather in its dynamical behavior and
particularly in the static, optical conductivity.  The electrical
conductivity, although static, is nevertheless a dynamical property,
since we need at least two different energies (small imaginary parts)
to determine it. The Kubo formula for the diagonal part of the
electrical conductivity in models with only elastically scattered
electrons can be written as\cite{Janis03a}
\begin{multline}\label{Conductivity-elastic}
  \sigma_{\alpha\alpha} = -\frac{e^2}{N^2}\sum_{\mathbf{k}\mathbf{k}'}
  v_\alpha(\mathbf{k}) v_\alpha(\mathbf{k}') \int_{-\infty}^\infty
  \frac{d\omega}{2\pi} \left\{\frac{df(\omega)}{d\omega}\right. \\
  \left.\left[
      G^{AR}_{\mathbf{k}\mathbf{k}'}(\omega,\omega;\mathbf{0})
      - G^{AA}_{\mathbf{k}\mathbf{k}'}(\omega,\omega;\mathbf{0}) \right]\right. \\
  \left.  + \frac 12 f(\omega)
    \frac{\partial}{\partial\omega}\left[G^{RR}_{\mathbf{k}
        \mathbf{k}'} (\omega,\omega;\mathbf{0}) -
      G^{AA}_{\mathbf{k}\mathbf{k}'}(\omega,\omega;\mathbf{0}) \right]
  \right\}\end{multline}%
where $v_\alpha(\mathbf{k}) = \partial\epsilon(\mathbf{k})/\partial
k_\alpha $ is the group velocity in the direction $\alpha$ and the
averaged (translationally invariant) two-particle Green functions are
defined
\begin{align*}
  G^{AR}_{\mathbf{k}\mathbf{k}'}(\omega,\omega';\mathbf{q}) &=
  G^{(2)}_{\mathbf{k} \mathbf{k}'}(\omega -i0^+, \omega' +
  i0^+;\mathbf{q}),\\
  G^{RR}_{\mathbf{k}\mathbf{k}'}(\omega,\omega';\mathbf{q}) &=
  G^{(2)}_{\mathbf{k} \mathbf{k}'}(\omega + i0^+, \omega' +
  i0^+;\mathbf{q})\ .
\end{align*}
We denoted $\omega = E- \mu$ the energy measured from the Fermi level
and $f(E) = 1/(1 + e^{\beta(E-\mu)})$ is the Fermi function. See
Fig.~\ref{fig:G2-Notation} for the way the variables in the
two-particle Green function are used in this paper.

The expression for the electrical conductivity simplifies if we resort
to zero temperature. Then the integrals over frequencies can be
performed explicitly and we obtain a simple formula
\begin{equation}\label{eq:Conductivity-zeroT}
  \sigma_{\alpha\alpha} = \frac{e^2}{2\pi N^2}\sum_{\mathbf{k}\mathbf{k}'}
  v_\alpha(\mathbf{k}) v_\alpha(\mathbf{k}')\left[ G^{AR}_{\mathbf{k}\mathbf{k}'}
    - \Re G^{RR}_{\mathbf{k}\mathbf{k}'}\right]
\end{equation}
with the values of the two-particle Green functions at the Fermi
energy. We used an abbreviation $G^{AR}_{\mathbf{k}\mathbf{k}'} =
G^{AR}_{\mathbf{k}\mathbf{k}'}(0,0;\mathbf{0})$.

The two-particle Green function $G^{(2)}$ carries information on both
the uncorrelated and correlated motion of two electrons. Only the
latter one is the genuine two-particle quantity. It is our task to
identify this contribution to the electrical conductivity. To do so,
we introduce the two-particle vertex $\Gamma$ defined from an equation
\begin{equation}\label{eq:G-Gamma}
  G^{AR}_{\mathbf{k}\mathbf{k}'} = G^A_\mathbf{k} G^R_\mathbf{k}\left[
    \delta(\mathbf{k} -\mathbf{k}') + \Gamma^{AR}_{\mathbf{k}\mathbf{k}'}
    G^A_{\mathbf{k}'} G^R_{\mathbf{k}'}\right]
\end{equation}
where we again used a notation $G^R_\mathbf{k} =
G^R_\mathbf{k}(0)$. With the aid of the vertex function we can
decompose the conductivity tensor into two parts
\begin{equation}\label{eq:Conductivity-decomposition}
  \sigma_{\alpha\alpha} = \sigma^{(0)}_{\alpha\alpha} + \Delta\sigma_{\alpha\alpha}
\end{equation}
where
\begin{equation}\label{eq:Conductivity-Drude}
  \sigma^{(0)}_{\alpha\alpha} = \frac{e^2}{\pi
    N}\sum_{\mathbf{k}}\left|v_\alpha(\mathbf{k}) \right|^2 \left|\Im
    G^R(\mathbf{k}) \right|^2
\end{equation}
is the standard one-electron or Drude conductivity at zero
temperature. The genuine two-particle contribution is called a vertex
correction and is proportional to the appropriate matrix element of
the two-particle vertex that at zero temperature reads
\begin{multline}\label{eq:Conductivity-correction}
  \Delta\sigma_{\alpha\alpha} = \frac{e^2}{2\pi
    N^2}\sum_{\mathbf{k}\mathbf{k}'} v_\alpha(\mathbf{k})
  v_\alpha(\mathbf{k}')\left\{
    \left|G^R_\mathbf{k}\right|^2\Delta\Gamma^{AR}_{\mathbf{k}\mathbf{k}'}
    \left| G^R_{\mathbf{k}'}\right|^2\right. \\ \left. -
    \Re\left[\left(G^R_\mathbf{k}\right)^2
      \Delta\Gamma^{RR}_{\mathbf{k}\mathbf{k}'}
      \left(G^R_{\mathbf{k}'}\right)^2 \right] \right\}\
  . \end{multline}
It is not the full two-particle vertex $\Gamma$ that is important for
the electrical conductivity, but only its odd part
$\Delta\Gamma$. That is, only the part of the vertex function being on
bipartite lattices an odd function in fermionic momenta $\mathbf{k}$
and $\mathbf{k}'$, contributes to the electrical conductivity.

\section{Equilibrium mean-field
  thermodynamics} \label{sec:Thermodynamics}

The Kubo formula for the electrical conductivity,
Eq.~\eqref{Conductivity-elastic}, was derived within the
linear-response theory and hence the Green functions entering this
formula are the equilibrium ones. To estimate quantitatively the
vertex corrections to the electrical conductivity we need to know the
equilibrium thermodynamics of FKM. To reduce the impact of
uncontrolled approximations we should at best know the exact
equilibrium grand potential. An exact solution to FKM is known in the
limit of infinite spatial dimensions (mean-field limit) with which we
start up. The equilibrium thermodynamics of the disordered FKM in
$d=\infty$ was analyzed in Ref.~\onlinecite{Janis92b}. The functional
of the averaged grand potential was found to be represented via a set
of complex variational parameters $G_n$ and $\Sigma_n$, where the
index $n$ corresponds to the $n$th fermionic Matsubara frequency
\begin{multline}\label{eq:FE-infty} \beta \left\langle
    \Omega\right\rangle_{av}\\ = - \sum_{n=-\infty}^\infty\left\{
    \int_{-\infty}^\infty d E \rho(E) \ln \left[i\omega_n + \mu - E -
      \Sigma_n\right]\right. \\ \left. + \left\langle \ln \left[1 +
        G_n(\Sigma_n - V) \right]\right\rangle_{av}\right.\bigg\} \\ -
  \left\langle \ln \left[1 + \exp\left\{\beta(\mu - E_f
        -\mathcal{E}_V\right\} \right]\right\rangle_{av}\
  . \end{multline}%
Symbol $\left\langle \ \right\rangle_{av}$ stands for averaging over
the distribution of the random potential $V_i$. The shift of the
$f$-electron atomic level $\mathcal{E}_V$ is determined via the same
complex numbers $G_n$ and $\Sigma_n$
\begin{equation}\label{eq:E-shift}
  \mathcal{E}_V = - T\sum_{n=-\infty}^\infty  \ln \left[1 - \frac {U
      G_n}{1 + G_n(\Sigma_n -V)}\right]\
\end{equation}
and depends on the configuration of the random atomic potential $V$.
 
The equilibrium thermodynamics is obtained as a stationarity point
with respect to small variations of complex numbers $\Sigma_n$ and
$G_n$ of the averaged grand potential $\left\langle
  \Omega\right\rangle_{av}$ from Eq.~\eqref{eq:FE-infty}.  Vanishing
of variations of the former and the latter parameters lead to a couple
of equations for each Matsubara frequency $i\omega_n$
\begin{subequations}\label{eq:1P-stationarity}
  \begin{align}\label{eq:1PGF}
    G_n &= \int_{-\infty}^\infty \frac{ d
      \epsilon\rho(\epsilon)}{i\omega_n + \mu - \epsilon -\Sigma_n}\
    ,\\ \label{eq:SE} 1 &= \left\langle \frac{1 - f(E_f +
        \mathcal{E}_V)}{1 + G_n(\Sigma_n - V)}+ \frac{f(E_f +
        \mathcal{E}_V)}{1 + G_n(\Sigma_n - V - U)}\right\rangle_{av} .
  \end{align} \end{subequations}%
The first equation states that in equilibrium $G_n$ is the local
element of the one-electron thermal Green function with a self-energy
$\Sigma_n$. The second equation determines the value of the
equilibrium self-energy. These equations of thermal equilibrium must
be completed with an equation determining the chemical potential $\mu$
from the total electron density $n$. This equation then is
\begin{align}\label{eq:n-def}
  n& = \left\langle f(E_f + \mathcal{E}_V)\right\rangle_{av} +
  \sum_{n=-\infty}^\infty G_n e^{i\omega_n0^+}\nonumber \\ &=
  \left\langle f(E_f + \mathcal{E}_V)\right\rangle_{av}\nonumber\\
  &\quad - \int_{-\infty}^\infty\!\! \frac{d\omega}\pi f(\omega) \Im
  G^R\left(\omega -\Sigma^R(\omega)\right)
\end{align}
where $G^R(z) = N^{-1}\sum_{\mathbf{k}}G^R(\mathbf{k},z)$.  Equations
\eqref{eq:E-shift}-\eqref{eq:n-def} fully determine the equilibrium
thermodynamics for a given temperature $T$ and a total particle
density $n$.

Only one-particle equilibrium functions can be directly calculated
from the grand potential $\left\langle \Omega\right\rangle_{av}$. To
derive higher-order correlation functions we have to slightly perturb
equilibrium and look at the corresponding response functions.  For the
electrical conductivity we need to know two-particle vertex
$\Gamma$. The only consistent way to derive a two-particle vertex
within the mean-field theory is to keep the non-equilibrium
perturbation local.\cite{Janis99} The resulting vertex remains local
and is only frequency (energy) dependent.

\begin{figure}
  \includegraphics[width=9cm]{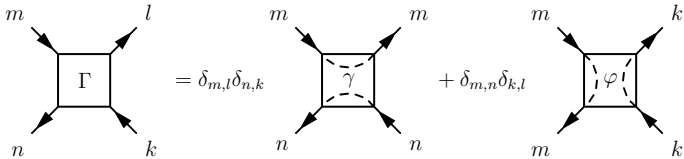}
  \caption{Graphical representation of Eq.~\eqref{eq:Gamma-local}. The
    dashed lines within the boxes indicate charge propagation from the
    incoming to the corresponding outgoing
    line. }  \label{fig:Gamma-Equation} \end{figure}

The equilibrium two-particle vertex has generally three independent
(Matsubara) frequencies. In the case of FKM investigated here, the
resulting two-particle vertex can have maximally two independent
frequencies.  The two frequencies can, however, be placed in two
different ways. The full local two-particle vertex for FKM can be
represented as
\begin{equation}\label{eq:Gamma-local}
  \Gamma^{MF}_{mn,kl} = \delta_{m,l} \delta_{n,k} \gamma_{m,n} + \delta_{m,n}
  \delta_{k,l}\varphi_{m,k}\ .
\end{equation}
The two contributions to the full vertex correspond to two ways the
electron lines go through the vertex. The electron line entering into
the vertex at the upper left corner goes out either via the upper
right corner (vertex $\gamma$) or via the lower left corner (vertex
$\varphi$). See Fig.~\ref{fig:Gamma-Equation} for a graphical
representation and notation used in Eq.~\eqref{eq:Gamma-local}. Due to
energy conservation in this model the incoming and outgoing
frequencies must equal. Notice that the two vertices $\gamma$ and
$\varphi$ never mix up and are completely detached in the
solution. The former vertex is relevant for the transport while the
latter one for the thermodynamics. It is hence sufficient to take into
account only vertex $\gamma$ for the calculation of the electrical
conductivity.

We can represent vertex $\gamma$ via a local Bethe-Salpeter equation
with a local irreducible vertex $\Lambda_{m,n}$. We have
\begin{equation}\label{eq:BS-local}
  \Gamma_{m,n} = \frac {\Lambda_{m,n}}{1 -\Lambda_{m,n}G_m G_n}\ .
\end{equation}
Charge conservation leads to a generalized Ward identity that matches
a non-equilibrium variation of the self-energy $\delta \Sigma_{m,n}$
with the equilibrium irreducible vertex $\Lambda_{m,n}$
\begin{equation}\label{eq:lambda-def}
  \Lambda_{m,n} = \frac{\delta\Sigma_{m,n}}{\delta G_{m,n}} = \frac 1{G_m
    G_n}\left[ 1 - \lambda^{-1}_{m,n}\right]
\end{equation}
where we abbreviated
\begin{multline}\label{eq:lambda-eq}
  \lambda_{m,n} = \left\langle \frac{1 - f(E_f +
      \mathcal{E}_V)}{\left[1 + G_m(\Sigma_m- V)\right]\left[1 +
        G_n(\Sigma_n - V))\right]} \right. \\ \left. +\ \frac{f(E_f +
      \mathcal{E}_V)}{\left[1 + G_m(\Sigma_m - V - U)\right]\left[1 +
        G_n(\Sigma_n - V - U)\right]}\right\rangle_{av}\
  . \end{multline}%
The latter representation was derived with the aid of stationarity
equations, Eqs.~\eqref{eq:1P-stationarity}.

We have derived all the necessary equilibrium quantities needed for
the calculation of the electrical conductivity.  We must, however, go
beyond the local, mean-field limit in order to calculate transport
properties. That is, we have to perturb the equilibrium with
time-dependent non-local excitations. To do it in a controlled manner
we use a perturbation expansion around a mean-field. This can be
consistently achieved within an asymptotic expansion in high spatial
dimensions.

\section{Expansion around mean field -- Vertex corrections to the
  electrical conductivity}
\label{sec:Around-MF}

An asymptotic expansion around a mean-field solution in $d=\infty$ for
non-interacting disordered electrons was recently derived by one of us
in Ref.~\onlinecite{Janis05a}. The only ingredients of such an
expansion are the local mean-field vertex $\gamma$ and the the
non-local one-electron propagator.  The relevance and applicability of
this expansion goes beyond the model of disordered electrons. It can
be applied to any model where energy is conserved in scattering events
(elastic scatterings) and where only spatial fluctuations matter.  The
disordered FKM studied here falls into this category and we can hence
use the expansion concept of Ref.~\onlinecite{Janis05a} for it. It is
a considerable advantage to expand around a mean-field solution, since
all the effects of scatterings on impurities are included already in
the local vertex $\gamma$. The variance of the potential fluctuations
is not a small parameter and the expansion terms do not depend on
whether the scatterings are due to thermally equilibrated or randomly
distributed static impurities, that is, whether the disorder is
annealed or quenched.

We construct the expansion around a mean field as an asymptotic series
on a hyper-cubic lattice in high spatial dimensions. The expansion
parameter is the off-diagonal one-electron propagator from the
mean-field theory. We define
\begin{equation}\label{eq:G-off}
  \bar{G}(\mathbf{k},\zeta) = \frac 1{\zeta - \epsilon(\mathbf{k})} - \int
  \frac{d\epsilon \rho(\epsilon)}{\zeta -\epsilon}
\end{equation}
where we denoted $\zeta= z - \Sigma(z)$ and the local self-energy
$\Sigma(z)$ is that of the mean-field solution. The off-diagonal two
particle bubble is a convolution of the off-diagonal one-electron
propagators. We hence define
\begin{equation}\label{eq:chi-bar}
  \bar{\chi}(\zeta,\zeta';\mathbf{q}) = \frac 1N\sum_{\mathbf{k}}
  \bar{G}(\mathbf{k},\zeta) \bar{G}(\mathbf{k} + \mathbf{q},\zeta')\ .
\end{equation}
The frequency indices are external parameters and we suppress them
when they are not necessary to specify a particular type of the one or
two-electron propagators.

The asymptotic limit of the full two-particle vertex in high spatial
dimensions contains beyond the local mean-field vertex $\gamma$ also
non-local contributions from the electron-hole and electron-electron
ladders.\cite{Janis05a} It can be represented as follows
\begin{equation}\label{eq:Gamma-high_d}
  \Gamma_{\mathbf{k}\mathbf{k}'}(\mathbf{q}) = \gamma\left[1 + \gamma
    \left(\frac{\bar{\chi}(\mathbf{q})}{1 - \gamma\bar{\chi}(\mathbf{q})}
      + \frac{\bar{\chi}(\mathbf{Q})}{1 - \gamma\bar{\chi}(\mathbf{Q})}\right)\right]
\end{equation}
where we denoted $\mathbf{Q} = \mathbf{q} + \mathbf{k} + \mathbf{k}'$
the momentum conserved in the electron-electron channel. Notice that
the contribution from the electron-hole channel with
$\bar{\chi}(\mathbf{q})$ is part of the two-particle vertex from the
Coherent Potential Approximation (CPA) and can be derived from a
Velick\'y-Ward identity.\cite{Velicky69} The two-particle vertex from
CPA does not carry the full $1/d$ correction to the local vertex and
moreover it is not electron-hole symmetric on the two-particle
level.\cite{Janis05a} A consistent extension of the local mean-field
two-particle vertex must contain both non-local contributions from the
electron-hole and the electron-electron channels as given in
Eq.~\eqref{eq:Gamma-high_d}.

The contribution from multiple scatterings in the electron-electron
channel to the asymptotic two-particle vertex is important in
particular in the calculation of the vertex corrections to the
mean-field electrical conductivity. The CPA vertex, as notoriously
known, does not generate vertex corrections, unless we introduce odd
dispersion relations in multi-orbital models. It is only the second
term in the parentheses on the right-hand side of
Eq.~\eqref{eq:Gamma-high_d}, that contributes to the averaged
conductivity and the we can identify
\begin{equation}\label{eq:Delta-Gamma}
  \Delta\Gamma_{\mathbf{k}\mathbf{k}'}(\mathbf{q}) =
  \gamma^2\frac{\bar{\chi}(\mathbf{k}+ \mathbf{k}' + \mathbf{q})}{1 -
    \gamma\bar{\chi}(\mathbf{k}+ \mathbf{k}' + \mathbf{q})}
\end{equation}
from the formula for the vertex corrections to the electrical
conductivity, Eq.~\eqref{eq:Conductivity-correction}. Inserting
representation from Eq.~\eqref{eq:Delta-Gamma} into
Eq.~\eqref{eq:Conductivity-correction} we obtain
\begin{multline}\label{eq:Delta-sigma}
  \Delta\sigma^{MF}_{\alpha\alpha} = \frac{e^2}{2\pi
    N^2}\sum_{\mathbf{k}\mathbf{k}'}\\\left\{
    \left(\gamma^{RA}\right)^2 \left|G^R_{\mathbf{k}}\right|^2
    \frac{v_\alpha(\mathbf{k}) \bar{\chi}^{RA}(\mathbf{k} +
      \mathbf{k}') v_\alpha(\mathbf{k}')}{1 - \gamma^{RA}
      \bar{\chi}^{RA}(\mathbf{k} +
      \mathbf{k}')}\left|G^R_{\mathbf{k}'}\right|^2 \right. \\
  \left. - \Re\left[\left(\gamma^{RR}\right)^2
      \left(G^R_{\mathbf{k}}\right)^2 \frac{v_\alpha(\mathbf{k})
        \bar{\chi}^{RR}(\mathbf{k} + \mathbf{k}')
        v_\alpha(\mathbf{k}')}{1 - \gamma^{RR}
        \bar{\chi}^{RR}(\mathbf{k} +
        \mathbf{k}')}\left(G^R_{\mathbf{k}'}\right)^2 \right]\right\}\
  .  \end{multline}%
It is a general formula for the leading-order corrections to the
mean-field (Drude) conductivity and can be applied in any dimension.
Vertex $\Delta\Gamma$ contains the so-called Cooper pole being the
image of the diffusion pole contained in the CPA vertex after an
appropriate electron-hole transformation. The Cooper pole is
responsible for the so-called weak-localization corrections in the
Anderson model of disordered electrons.\cite{Akkermans07} These
corrections are here identical with those determined by
Eq.~\eqref{eq:Delta-sigma}. They are negative and diverge in spatial
dimensions $d\le 2$. The mean-field corrections to the electrical
conductivity from Eq.~\eqref{eq:Delta-sigma} can effectively be
applied only in dimensions $d\ge 3$ and not too close to the band
edges.

We can further simplify Eq.~\eqref{eq:Delta-sigma} in that we get rid
of the local mean-field vertex $\gamma$. We utilize the Ward identity
from Eq.~\eqref{eq:lambda-def} connecting the one-electron self-energy
$\Sigma$ and the two-particle irreducible vertex $\Lambda$. Using the
stationarity equation, Eq.~\eqref{eq:1PGF}, we easily obtain
\begin{subequations}\label{eq:WI}
  \begin{align}
    \Lambda_{+-} &= \frac{\Im\Sigma_{+}}{\Im G_{+}} = \frac 1
    {{\chi}_{+-}(\mathbf{0})}\ , \\ \Lambda_{++} &=
    \frac{\Sigma_{+}^\prime}{ G_{+}^\prime} = \frac {Z_+}
    {{\chi}_{++}(\mathbf{0})}\ \end{align}\end{subequations}%
where we denoted $\Sigma_+^\prime
= \partial\Sigma_+(\omega)/\partial\omega|_{\omega=0}$, $G_+^\prime
= \partial G_+(\omega)/\partial\omega|_{\omega=0}$, and $Z_+
=\Sigma_+^\prime /(\Sigma_+^\prime - 1) $.  With these relations and
the local Bethe-Salpeter equation, Eq.~\eqref{eq:BS-local}, we can
rewrite Eq.~\eqref{eq:Delta-sigma} to
\begin{widetext}
  \begin{multline}\label{eq:Delta-sigma-WI}
    \Delta\sigma^{MF}_{\alpha\alpha} = \frac{e^2}{2\pi
      N^2}\sum_{\mathbf{k}\mathbf{k}'} \left\{
      \frac{\left|G^R_{\mathbf{k}}\right|^2 v_\alpha(\mathbf{k})
        \bar{\chi}_{+-}(\mathbf{k} + \mathbf{k}')
        v_\alpha(\mathbf{k}')\left|G^R_{\mathbf{k}'}\right|^2}
      {\left(\chi_{+-}(\mathbf{0}) - \left|\langle G_+\rangle\right|^2
        \right) \left(\chi_{+-}(\mathbf{0}) - \chi_{+-}(\mathbf{k} +
          \mathbf{k}')\right)} \right. \\ \left. - \Re\left[
        \frac{Z_+^2\left(G^R_{\mathbf{k}}\right)^2
          v_\alpha(\mathbf{k}) \bar{\chi}_{++}(\mathbf{k} +
          \mathbf{k}') v_\alpha(\mathbf{k}')
          \left(G^R_{\mathbf{k}'}\right)^2}
        {\left(\chi_{++}(\mathbf{0}) - Z_+ \langle G_+\rangle^2
          \right) \left(\chi_{++}(\mathbf{0}) -
            Z_+\chi_{++}(\mathbf{k} + \mathbf{k}')\right)}
      \right]\right\}
  \end{multline}
\end{widetext}
where we used an abbreviated notation
\begin{subequations}\label{eq:G-linear}
  \begin{align}
    \left\langle G_\pm\right\rangle &= \int_{-\infty}^\infty \frac
    {d\epsilon
      \rho(\epsilon)}{\left[E_F - \epsilon - \Sigma_\pm \pm i0^+ \right]}\ ,\\
    \label{eq:G-square}
    \left\langle G_\pm^2\right\rangle &= \int_{-\infty}^\infty \frac
    {d\epsilon \rho(\epsilon)}{\left[E_F - \epsilon - \Sigma_\pm \pm
        i0^+ \right]^2}\ .
  \end{align}\end{subequations}
Representation \eqref{eq:Delta-sigma-WI} of the vertex corrections to
the electrical conductivity does not explicitly contain the strength
of elastic scatterings in the model.  This strength is beyond the
self-energy of the one-electron propagators implicitly comprised in
the spatial fluctuations of the two-particle bubble
$\chi(\mathbf{q})$. A singular structure of the integrand in momentum
representation of the vertex corrections to the Drude conductivity
becomes transparent in Eq.~\eqref{eq:Delta-sigma-WI}.

\section{Vertex corrections to the electrical conductivity from high
  spatial dimensions}
\label{sec:Leading-corrections}

To evaluate the vertex corrections to the mean-field electrical
conductivity, Eq.~\eqref{eq:Delta-sigma}, we resort to high spatial
dimensions where we can explicitly perform the convolutions over
momenta. We are interested only in the leading-order contributions in
the inverse spatial dimension $1/d$. The conductivity vanishes in the
mean-field limit, $d=\infty$, where only local quantities survive. The
actual mean-field conductivity is due to the velocity in a particular
direction proportional to $1/d$ and hence can be treated only
asymptotically for $d\to\infty$. On a hyper-cubic lattice we have
$v_\alpha(\mathbf{k}) = \partial \epsilon(\mathbf{k})/\partial
k_\alpha = t d^{-1/2} \sin k_\alpha$.  Inserting this result in the
mean-field conductivity, Eq.~\eqref{eq:Conductivity-Drude}, we obtain
the Drude conductivity on a hyper-cubic $d$-dimensional lattice to be
\begin{align}\label{eq:sigma0-highd}
  \sigma_0 & = \frac{e^2t^2}{4 \pi d }\left[\left\langle|
      G_+|^2\right\rangle -
    \Re\left\langle G_+^2\right\rangle\right] \nonumber \\
  & = \frac{e^2t^2}{2 \pi d } \Im\Sigma^2 \left\langle|
    G_+G_-|^2\right\rangle \ . \end{align}

To evaluate the vertex corrections we represent the one and
two-particle propagators so that we can separate the Cartesian
components of momenta. We use the following integral representation
for the one-electron propagator
\begin{equation}\label{eq:G-integral}
  G(\mathbf{k},\zeta_\pm) = -i \int_{0}^{\infty} d u\ e^{\pm i u\zeta_\pm}
  \prod_{\nu =1}^d \exp\{\pm \frac {itu}{\sqrt{d}}\cos k_\nu \} \ ,
\end{equation}
where we assumed that $\Im\zeta_+>0$ and $\Im\zeta_-<0$ in order to
keep the integrals convergent. Here $k_\nu$ is the $\nu$th Cartesian
component of momentum $\mathbf{k}$ on a $d$-dimensional hyper-cubic
lattice.

The two-particle bubble can be represented in a similar
way. Performing the integration over momenta we obtain in the leading
asymptotic order for $d\to\infty$
\begin{multline}\label{eq:chi-int}
  \chi(\mathbf{q};\zeta,\zeta') = -\int_0^\infty du \int_0^\infty d v\
  e^{i u \zeta} e^{iv\zeta'}\\ \exp\{\frac{t^2(u^2 +
    v^2)}4\}\prod_{\nu=1}^d \exp\{- \frac{uvt^2}{2d}\cos q_\nu \} \ .
\end{multline}%
For simplicity we assumed that both complex energies $\zeta$ and
$\zeta'$ have positive imaginary parts. A generalization to different
imaginary parts is straightforward.

The above integral representations suffice to evaluate the leading
order of the vertex corrections. We realize that the denominator in
the representation of the vertex corrections,
Eq.~\eqref{eq:Delta-sigma}, does not contribute in the leading order
for $d\to\infty$. We denote
\begin{equation}\label{eq:J-integral}
  J^{RR}_{\alpha\alpha} = \frac 1{N^2}\sum_{\mathbf{k}\mathbf{k}'}
  \left(G^R_{\mathbf{k}}\right)^2 v_\alpha(\mathbf{k})
  \bar{\chi}^{RR}(\mathbf{k} + \mathbf{k}') v_\alpha(\mathbf{k}')
  \left(G^R_{\mathbf{k}'}\right)^2\ .
\end{equation}
We use the integral representation separating the Cartesian components
of momenta. We then have
\begin{multline*}
  J^{RR}_{\alpha\alpha} = \frac{t^2}d \int_0^\infty d a \int_0^\infty
  d b \int_0^\infty d a'  \int_0^\infty d b'   \int_0^\infty d u \\
  \int_0^\infty d v \ e^{i(a+b+a'+b'+u+v)\zeta_+}
  \exp\left\{-\frac{t^2(u^2 + v^2)}{4}\right\}\\ \times \sin k_\alpha
  \sin k'_\alpha \prod_{\nu=1}^d \exp\left\{\frac{i(a +
      b)t}{\sqrt{d}}\cos k_\nu \right. \\ \left.+\ \frac{i(a' +
      b')t}{\sqrt{d}}\cos k'_\nu - \frac{uvt^2}{2d}\cos(k_\nu +
    k'_\nu) \right\}\ . \end{multline*}
The leading non-vanishing contribution from the summation over momenta
is proportional to $1/d$. Performing the calculation we obtain
\begin{multline}\label{eq:J-asymptotic}
  J^{RR}_{\alpha\alpha} = \frac{t^4}{8d^2} \int_0^\infty d a
  \int_0^\infty d b \int_0^\infty d a'  \int_0^\infty d b' \int_0^\infty d u\\
  \int_0^\infty d v\ e^{i(a+b)\zeta_+}\exp\left\{-\frac{(a +
      b)^2t^2}{4} \right\} e^{i(a'+b')\zeta_+}\\ \times
  \exp\left\{-\frac{(a' + b')^2t^2}{4} \right\} e^{iu\zeta_+}
  \exp\left\{\frac{u^2t^2}4 \right\} \\ \times e^{iv\zeta_+}\
  \exp\left\{\frac{v^2t^2}4 \right\} u v\ . \end{multline}
Using representation \eqref{eq:G-linear} we can rewrite
$J^{RR}_{\alpha\alpha}$ in a simple form
\begin{subequations}\label{eq:J-simple}
  \begin{equation}\label{eq:J-RR}
    J^{RR}_{\alpha\alpha} = - \frac{t^4}{8d^2} \left\langle G_+ G_+\right\rangle^2
    \left\langle G_+^2\right\rangle \left\langle G_+^2\right\rangle
  \end{equation}
  where each Green function $G_+$ stands for the selection of the
  imaginary part of the propagator, the retarded one. In the case of
  the retarded and advanced propagators we then have
  \begin{equation}\label{eq:J-RA}
    J^{RA}_{\alpha\alpha} = - \frac{t^4}{8d^2} \left\langle G_+ G_-\right\rangle^2
    \left\langle G_+^2\right\rangle \left\langle G_-^2\right\rangle \ .
  \end{equation}\end{subequations}
We use these results to represent the vertex correction to the
electrical conductivity in high spatial dimensions. We obtain
\begin{multline}\label{eq:Delta-sigma1}
  \Delta_d\sigma = - \frac{e^2t^4}{16\pi d^2} \left\{\left\langle
      \left|G_+\right|^2\right\rangle^2 \left|\left\langle
        G_+^2\right\rangle\right|^2 \gamma_{+-}^2 \right. \\ \left. -
    \Re\left[\left\langle G_+^2\right\rangle^4 \gamma_{++}^2 \right]
  \right\}\ . \end{multline}
With the aid of the Ward identity, Eq.~\eqref{eq:WI}, we can further
simplify the expression for the leading contribution to the electrical
conductivity beyond the Drude formula in $d$ spatial dimensions
\begin{multline}\label{eq:Delta-sigma2} \Delta_d\sigma = -
  \frac{e^2t^4}{16\pi d^2} \left\{\frac{\left\langle
        \left|G_+\right|^2\right\rangle^2\left|\left\langle
          G_+^2\right\rangle\right|^2} {\left[\left\langle
          \left|G_+\right|^2\right\rangle - \left| \left\langle
            G_+\right\rangle \right|^2\right]^2} \right.  \\ \left. -
    \Re \left[\frac{Z_+^2\left\langle G_+^2\right\rangle^4}
      {\left[\left\langle G_+^2\right\rangle - Z_+ \left\langle
            G_+\right\rangle^2 \right]^2 } \right]\right\}\
  .\end{multline}

The sign of the vertex correction is negative. Although it is only the
leading asymptotic term, it determines the overall sign of the vertex
correction. The neglected higher-order terms must be summed to
infinite order so that the denominators on the right-hand side of
Eq.~\eqref{eq:Delta-sigma-WI} are recovered. The higher-order terms
then do not change the sign of the leading-order vertex correction to
the electrical conductivity.

\section{Results}\label{sec:Reults}

To reach numerical values for the vertex corrections to the electrical
conductivity we need the stationarity equation for the self-energy,
Eq.~\eqref{eq:SE}. This equation significantly simplifies in the pure
model at half filling with a symmetry between $c$ and $f$ electrons,
it means if $n_c=1/2$, $n_f=1/2$, $E_f=0$, and $E_F=U/2$.  We then
obtain $\Re\Sigma(0) = U/2$ and the solution for the local functions
can be expressed in terms of a single parameter
$$x=\langle \left|G_+\right|^2\rangle = \int_{-\infty}^\infty  \frac{d\epsilon
  \rho(\epsilon)}{\epsilon^2 + \Im\Sigma_+^2} >0\ .$$
It is easy to find from Eq.~\eqref{eq:SE} that
\begin{subequations}\label{eq:1P-EHS}
  \begin{align}\label{eq:Sigma-EHS}
    \Im\Sigma_+ &= - \sqrt{\frac 1x - \frac {U^2}4}\\
    \label{eq:G-EHS}
    \Im G_+ & =- \sqrt{x\left( 1 - \frac {U^2}4 x\right)}\\
    \label{eq:gammapm-EHS}
    \gamma_{+-} & = \frac {4}{U^2 x^2} = \gamma_{++} & \ .
  \end{align}\end{subequations}%
The stationarity equation for the parameter $x$ reads
\begin{equation}
  \label{eq:x-EHS}
  1  = \int_{-\infty}^\infty  \frac{d\epsilon \rho(\epsilon)} {x\left(\epsilon^2 -
      \frac {U^2}4 \right) + 1} \equiv \left\langle \frac 1{x\left(\epsilon^2 - \frac
        {U^2}4 \right) + 1} \right\rangle
\end{equation}
where we again abbreviated the integral over energy weighted by the
density of energy states by angular brackets.

The two-electron bubble for zero momentum can be represented as
follows
\begin{subequations}\label{eq:2P-EHS}
  \begin{align}
    \label{eq:G2-EHS} \left\langle G_+^2\right\rangle & = - x \left[ 1
      - \frac {U^2}2 x + 2 \left(1 - \frac{U^2}4 x
      \right)x_2\right], \end{align}
  where we introduced variance of non-local fluctuations
  \begin{align}
    x_2 & = \frac 1{x^2} \left[\int_{-\infty}^\infty \frac{d\epsilon
        \rho(\epsilon)}{\left[\epsilon^2 + \Im\Sigma_+^2\right]^2} -
      x^2 \right] \ .
  \end{align}
\end{subequations}%

The Drude conductivity then is
\begin{equation}\label{sigma0:x-EHS}
  \sigma_0 = \frac{e^2}{2\pi d}\ x\left(1 - \frac {U^2}4 x \right)\left(
    1 + x_2 \right)
\end{equation}
and the vertex correction reads
\begin{multline}\label{eq:Dsigma-EHS}
  \Delta_d\sigma \\ =- \frac{ e^2 }{\pi d^2}\ \frac {1 - \frac {U^2}4
    x}{U^4}\left[ 1 - \frac {U^2}2 x + 2 \left(1 - \frac{U^2}4 x
    \right)x_2\right]^2\\ \times \left\{U^2 x - 4 x_2\left[1 -
      \frac{U^2}2 x + \left(1 - \frac{U^2}4 x\right)x_2\right]\right\}
  \ .
\end{multline}%
We set the energy scale $t=1$.  Note that $0\le x \le 4/U^2$ for $U^2
\le 4\left\langle \epsilon^2\right\rangle$. The upper limit on the
interaction strength $U$ in the above equations is imposed by the
metal-insulator transition at which $x=0$ and the density of the
extended electrons at the Fermi level vanishes.

The Drude conductivity in the weak-coupling limit ($U\to 0$) diverges,
that is the resistivity vanishes as it should be for the Fermi gas
without impurity scatterings.  The two parameters $x$ and $x_2$ behave
in the weak-coupling limit $x\sim 4/U^2 - \pi^2\rho(0)^2$ and $x_2
\sim 2/U^2\pi^2\rho(0)^2 - 1$ and hence
\begin{align}\label{eq:U0-Drude}
  \sigma_0 & = \frac{e^2}{\pi d}\ \frac 1{U^2}\ .
\end{align}
The vertex correction from Eq.~\eqref{eq:Dsigma-EHS} remains finite in
the weak-coupling limit. To derive an explicit expression for it we
had to expand the asymptotic solution for the parameters $x$ and $x_2$
up to the third order in $d^{-1}$.

\begin{figure}
  \includegraphics[width=7cm]{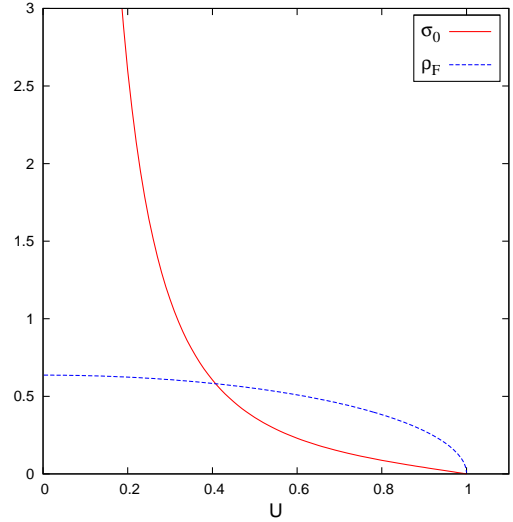}
  \caption{(Color online) Drude conductivity vs. density of states for
    the electron-hole symmetric Falicov-Kimball model with
    semi-elliptic density of states and
    $d=3$. }  \label{fig:S0_EHS} \end{figure}

\begin{figure}
  \includegraphics[width=7cm]{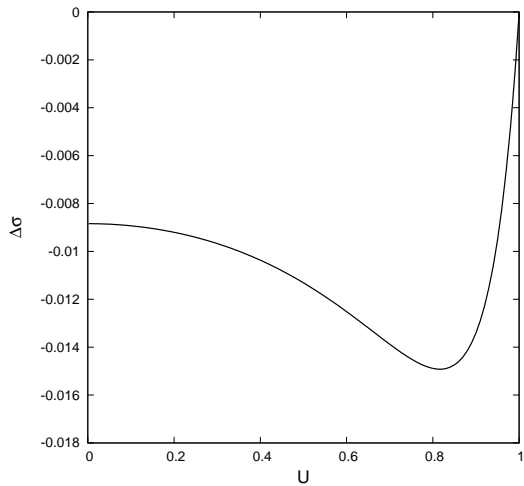}
  \caption{Vertex correction to the conductivity for the electron-hole
    symmetric case.  }\label{fig:DS_EHS} \end{figure}

\begin{figure}
  \includegraphics[width=7cm]{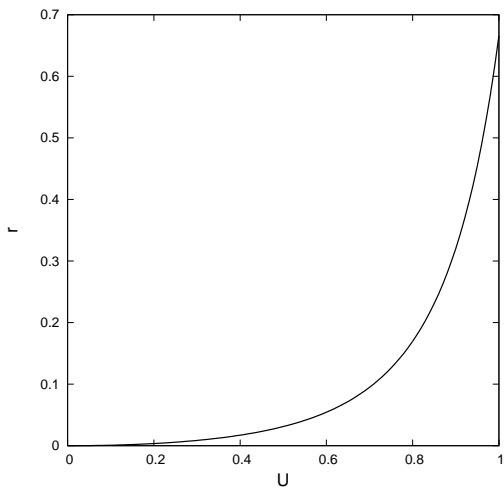}
  \caption{Ratio $r=|\Delta\sigma|/\sigma_0$ for the electron-hole
    symmetric case. }\label{fig:r_EHS} \end{figure}

For the explicit calculation we used the semi-elliptic density of
states $\rho(\omega) = 2/\pi\sqrt{1 - \omega^2}$. We choose the
dimensionality parameter $d=3$ in our calculations. The Drude
conductivity is plotted in Fig.~\ref{fig:S0_EHS}. It is compared to
the density of states (DOS) of the mobile electrons at the Fermi
energy. Coulomb interaction decreases both the DOS and the
conductivity down to the metal-insulator transition, where the
self-energy diverges. The vertex correction $\Delta\sigma$ from
Eq.~\eqref{eq:Dsigma-EHS} is plotted in Fig.~\ref{fig:DS_EHS}. The
modulus of the vertex correction is not a monotonic function and it
reaches maximum at about $U_m\approx 0.82$. The vertex correction is
negative but is much smaller than the Drude conductivity. Their ratio
is plotted in Fig.~\ref {fig:r_EHS}. Only close to the metal-insulator
transition the vertex correction is of order of the mean-field
conductivity.

\begin{figure}
  \includegraphics[width=7cm]{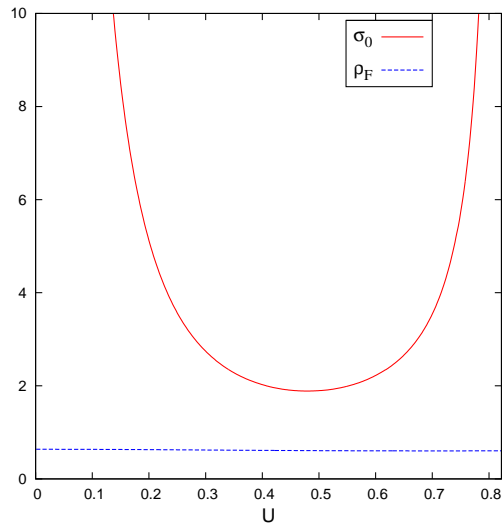}
  \caption{(Color online) Drude conductivity vs. density of states for
    filling $n=0.7$.  } \label{fig:S0_07}\end{figure}

\begin{figure}
  \includegraphics[width=7cm]{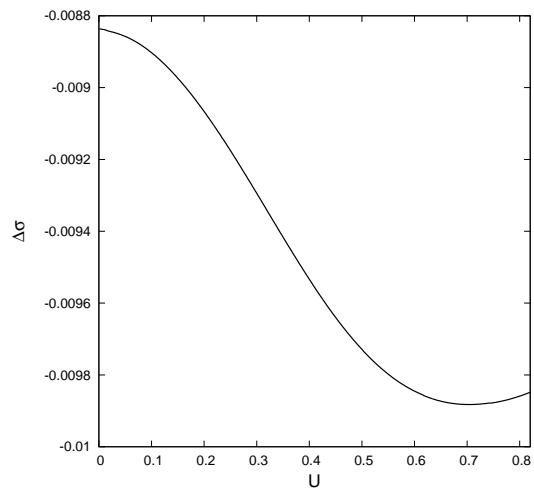}
  \caption{Vertex correction to the conductivity for filling $n=0.7$.
  } \label{fig:DS_07} \end{figure}

\begin{figure}
  \includegraphics[width=7cm]{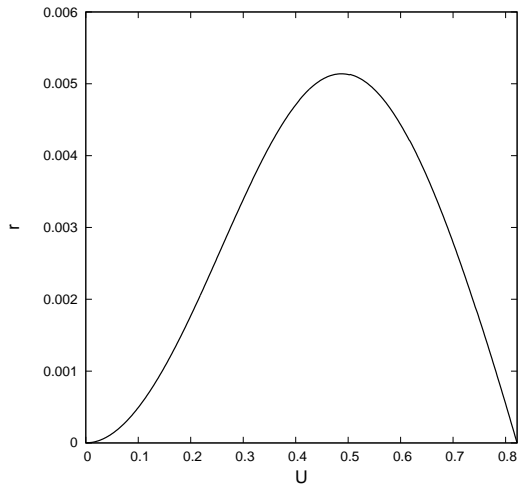}
  \caption{ Ratio $r=|\Delta\sigma|/\sigma_0$ for filling
    $n=0.7$. } \label{fig:r_07} \end{figure}

The Falicov-Kimball model away from half filling is more complicated.
First, we need two parameters to describe the local quantities.  They
are
$$x=\langle \left|G_+\right|^2\rangle = \int_{-\infty}^\infty  \frac{d\epsilon
  \rho(\epsilon)}{\left(E_F - \Re\Sigma -\epsilon\right)^2 +
  \Im\Sigma_+^2} $$ and
$$
y = \int_{-\infty}^\infty \frac{d\epsilon
  \rho(\epsilon)\epsilon}{\left(E_F - \Re\Sigma -\epsilon\right)^2 +
  \Im\Sigma_+^2}
$$
where we subtract the Fermi energy and the self-energy from their
values in the electron-hole symmetric case $n_f=1/2,
n_c=1/2$. Explicit formulas are more involved and we do not present
them here.  Second, as discussed in Ref.~\onlinecite{Si92}, a
nontrivial solution at zero temperature exists only for $1/2 < n <
3/2$. Outside this region the $f$-electron energy level is either
empty ($n\le 1/2$) or fully filled ($n\ge 3/2$).  We plotted the Drude
conductivity and DOS in Fig.~\ref{fig:S0_07} for a total filling
$n=0.7$. The density of states at the Fermi energy is almost constant
within the interval $0\le U\le 0.82$, within which the extended
electrons scatter on $f$-electrons. The extended electrons go over to
a Fermi gas in both limiting values of interaction, hence the Drude
conductivity diverges at both ends. The asymptotics is, however,
different at the two ends.  The asymmetry between the two limiting
interaction strengths is demonstrated in Fig.~\ref{fig:DS_07}, where
we plotted the vertex correction to the Drude conductivity. It is two
orders smaller than the mean-field one. This is clearly seen from
their ratio plotted in Fig.~\ref{fig:r_07}. The ratio is no longer a
monotonically increasing function as in the electron-hole symmetric
case, but reaches maximum at $U_m\approx 0.48$.

\section{Conclusions}
\label{sec:Conclusions}

We studied effects of elastic scatterings of mobile electrons on
either thermally equilibrated or frozen, randomly distributed static
impurities. We concentrated on quantum coherence effects due to
correlated back-scatterings and its impact on the electrical
conductivity. We calculated vertex corrections to the mean-field
(Drude) conductivity. We used a systematic expansion around the
mean-field solution via the asymptotic limit to high spatial
dimensions. Our principal finding is that elastic scatterings always
lead to diminution of the Drude conductivity. That is, vertex
corrections due to elastic scatterings have negative sign. The sign of
the vertex correction is determined by its leading high-dimensional
term, that is by the numerators in Eq.~\eqref{eq:Delta-sigma-WI}.  The
sign of the vertex correction is not affected by the type of
randomness in the distribution of the scattering impurities.  Coulomb
interaction in the pure Falicov-Kimaball model has the same effect on
the electrical conductivity as variance of fluctuations of the atomic
potential in the disordered Anderson model, it hinders diffusion.
 
We explicitly calculated only the leading high-dimensional vertex
correction to the Drude zero-temperature conductivity in the
paramagnetic phase of FKM. Quantitatively the correction is almost
everywhere two orders smaller than the Drude term. It is due to the
fact that the Drude conductivity diverges in the limit of the Fermi
gas while the vertex correction asymptotically approaches a finite
value when the interaction is switched off. Only close to the
metal-insulator transition in the electron-hole symmetric case the
vertex correction is of order of the Drude one, however, they both
vanish at the transition point. The leading high-dimensional
contribution to the vertex correction of the mean-field conductivity
is quantitatively negligible. It is important only for determining the
sign of the vertex correction. To obtain a more realistic values of
the vertex corrections in low-dimensional systems, one has to consider
the full representation of the vertex correction from
Eq.~\eqref{eq:Delta-sigma-WI} containing the Cooper pole. Only then we
are able to include a sizable impact of spatial dimensionality.

The method for evaluating vertex corrections to the mean-field
electrical conductivity via an asymptotic expansion in high spatial
dimensions is universal and is suitable for any model with elastically
scattered electrons. We can hence use it to investigate whether the
Coulomb interaction in FKM can lead to suppression of diffusion in
low-dimensional systems as well as to study an interplay between the
Anderson localization and Mott-Hubbard metal-insulator transition in
the disordered Falicov-Kimball model.

\section*{Acknowledgment} %

Research on this problem was carried out within project AV0Z10100520
of the Academy of Sciences of the Czech Republic and supported in part
by Grant No. 202/07/0644 of the Grant Agency of the Czech Republic. We
thank M. \v{Z}onda for illuminating discussions on the numerical
solution of the Falicov-Kimball model and J. Koloren\v{c} for
discussions on behavior of the Drude conductivity in the weak
scattering limit.

 \end{document}